\documentclass[aps,prb,reprint,amsmath,amssymb,floatfix,longbibliography]{revtex4-2}

\usepackage{graphicx}
\usepackage{amsmath,amssymb,bbold,bm,color}
\usepackage{siunitx}
\usepackage{float}
\usepackage{epstopdf}
\usepackage{hyperref}
\usepackage[dvipsnames]{xcolor}
\usepackage[normalem]{ulem}

\usepackage{comment}
\usepackage{color, colortbl}
\definecolor{LightCyan}{rgb}{0.88,1,1}

\usepackage[resetlabels,labeled]{multibib}

\newcommand{\bee}{\begin{equation}}
\newcommand{\ee}{\end{equation}}

\newcommand{\note}[1]{{\color{blue}#1}}
\newcommand{\warn}[1]{ \textbf{\color{red}#1}}

\hypersetup{colorlinks=true,linkcolor=blue,citecolor=blue,urlcolor=blue}

\begin{document}

\title{NMon: enhanced transmon qubit based on parallel arrays of Josephson junctions}

\author{Oguzhan Can}
\thanks{Corresponding author: ocan@phas.ubc.ca.} 
\affiliation{Quantum Matter Institute and Department of Physics and Astronomy, University of British Columbia, Vancouver BC, Canada V6T 1Z4} 

\author{Marcel Franz}
\affiliation{Quantum Matter Institute and Department of Physics and Astronomy, University of British Columbia, Vancouver BC, Canada V6T 1Z4}

\date{\today}

%
\begin{abstract}
We introduce a novel superconducting qubit architecture utilizing parallel arrays of Josephson junctions. This design offers a substantialy improved relative anharmonicity, typically within the range of $|\alpha_r| \approx 0.1 - 0.3$, while maintaining transition matrix elements in both the charge and flux channels that are on par with those of transmon qubits. Our proposed device also features exceptional tunability and includes a parameter regime akin to an enhanced version of the fluxonium qubit. Notably, it enables an additional order of magnitude reduction in matrix elements influenced by flux noise, thus further enhancing its suitability for quantum information processing applications.

\end{abstract}

\date{\today}
\maketitle

\section{Introduction}
To effectively process quantum information, we must first learn how to store and access it reliably. The aim is to create a memory element that is not only sufficiently isolated to prevent data loss from environmental interference, but also constructed in a way that it is easy to access as needed. One popular approach to this problem has been to store quantum information in the energy levels of an artificial anharmonic oscillator which can be engineered using superconducting circuit elements \cite{ Krantz2019}, most notably the Josephson junctions. Of course, there are many other approaches to quantum information processing in different physical platforms which include quantum dots \cite{Chatterjee2021}, Rydberg atoms \cite{Saffman2016}, NV centers \cite{Pezzagna2021} and photonic circuits \cite{Wang2020}. 

Transmon~\cite{Koch2007} is one of the most extensively studied superconducting qubits, where the Josephson energy $E_J$ and the charging energy $E_C$ are chosen such that $E_J/E_C \gg 1$. This limit suppresses the noise due to charge fluctuations, which was a major challenge for the Cooper pair box \cite{Vion2002}, operated in the regime where $E_J \approx E_C$.  On the other hand, approaching the transmon limit also reduces the energy level anharmonicity measured by the relative anharmonicity parameter $\alpha_r$, a crucial property of the energy spectrum that prevents information from leaking out the computational space~\cite{Koch2007}. Therefore $E_J$ can not be made arbitrarily larger than $E_C$ and is typically chosen such that $E_J/E_C \approx 100$. Larger anharmonicity is desirable not only for separating the computational space from other energy levels, but also for decreasing the lower bound on the pulse duration $\tau_p \approx|\omega_{01}\alpha_r|^{-1}$ for the manipulation of the two level system, leading to faster gate operations \cite{Koch2007}. 

As an alternative to transmon, the flux qubit \cite{Orlando1999} solves the anharmonicity problem by introducing three junctions in a ring geometry, creating a double-well potential where the two lowest lying energy levels are energetically well separated from higher excited states. However, such geometries with loops are susceptible to flux fluctuations \cite{Yan2016}, reducing coherence. Fluxonium qubit \cite{Manucharyan2009} alleviates this problem by shunting a Josephson junction by an array of $N$ Josephson junctions where the flux fluctuations are supressed by a factor of $\sim 1/N$.

\begin{figure}[b]
	\centering
	\includegraphics[width=0.65\columnwidth]{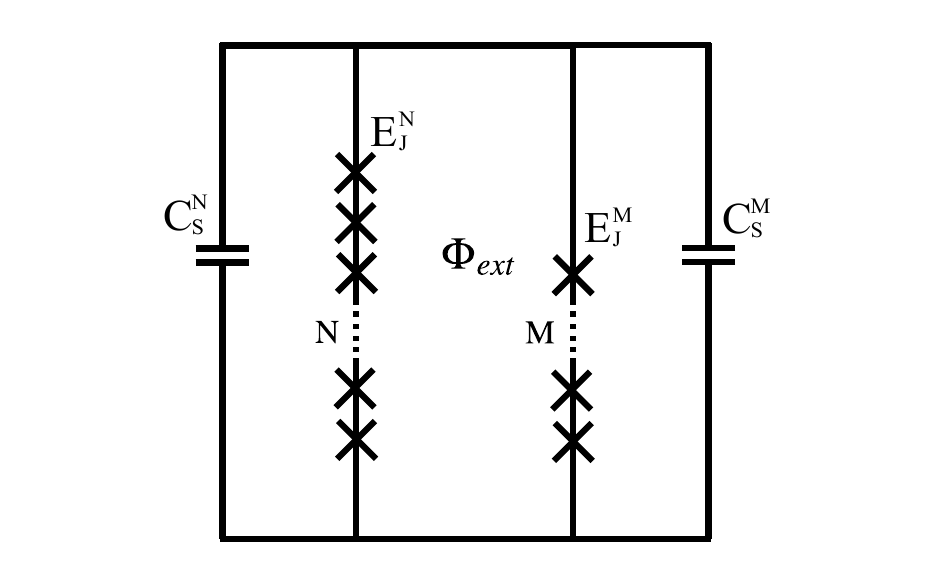}
	\caption{NMon geometry. Two parallel arms with $N$ and $M$ Josephson junctions shunted by large capacitances $C_S^N$ and $C_S^M$, respectively. The external flux $\Phi_{\rm ext}$ through the circuit loop introduces a relative phase difference between the harmonics contributing to the Josephson potential given in Eq.~\eqref{eq:nmon_intro}}.
	\label{fig:nmon}
\end{figure}

In this work we introduce the NMon with architecture shown in Fig.\ \ref{fig:nmon}, which can be thought of as a variant on split transmon for a certain choice of parameters. As we shall demonstrate, NMon exhibits enhanced anharmonicity compared to the transmon while maintaining transition matrix elements comparable in magnitude to those of the split transmon. We also show that if we increase the number of junctions in NMon, it has the following advantage over the fluxonium qubit: for the same number of Josephson junctions, while the matrix elements for charge fluctuations are comparable for the two systems, NMon transition matrix elements for flux fluctuations can be made smaller by an order of magnitude compared to the fluxonium. We expect NMon to work best near zero external flux through the loop, which happens to be the sweet spot where the energy levels show minimum dispersion as a function of the external flux. This is in contrast to the fluxonium  which is normally operated at half magnetic flux quantum bias, requiring currents driven through nearby flux bias loops. This tends to complicate the design and make scalability more challenging.

Our perspective in this paper is the free energy potential engineering using arrays of Josephson junctions for enhanced tunability and higher fidelity. The phase dynamics in the device depicted in Fig.\ \ref{fig:nmon} is described by the Hamiltonian
\begin{multline}
\label{eq:nmon_intro}
    H = 4E_C\hat{n}^2 -NE_J^N\cos \left(M\phi -\kappa \frac{\varphi_{\rm ext}}{N}\right) \\ -ME_J^M\cos \left(N\phi  + (1-\kappa)\frac{\varphi_{\rm ext}}{M}\right) 
\end{multline} 
where we defined $0 < \kappa < 1$ as $\kappa = C_S^M/(C_S^N+C_S^M)$, a parameter controlled by the ratio of the shunt capacitances for each arm and the effective charging energy
\begin{equation}
  E_C = \frac{2e^2}{4(NM)^2(C_S^M + C_S^N)}.
\end{equation}

Hamiltonian Eq.~\eqref{eq:nmon_intro} is derived in Appendix A using the method of canonical quantization with irrotational constraints \cite{You2019,Bryon2023} which is instrumental since we will consider time dependent flux  as a noise source and drive. If the external flux $\Phi_{\rm ext} = \varphi_{\rm ext} \frac{\Phi_0}{2\pi}$ through the loop of the circuit is time independent, we are free to allocate it in either cosine term when we impose the continuity condition for the superconducting order parameter. This can be considered a gauge choice. However, the time dependent case is more subtle and, as argued in Ref.\ \cite{You2019}, must be treated by imposing the irrotational constraints. In contrast to time independent flux case, different allocations of the flux controlled by the parameter $\kappa$ correspond to different values of junction capacitances or shunt capacitances in a specific limit which is described in detail in Appendix A. 

In addition to the external parameters such as flux and charge offset, the energy spectrum of NMon is determined by two dimensionless quantities 
\begin{equation}
  \beta=E_J^N/E_C, \ \ \ \ \eta=E_J^M/E_C.
\end{equation}
These are obviously invariant under rescaling of all energies which we will use to match the qubit frequency to a desired value.

\section{A variant on split transmon with enhanced anharmonicity}
Energy spectrum of transmon is controlled by the ratio $E_J/E_C$. As we increase this ratio the system becomes exponentially less susceptible to charge noise but at the same time anharmonicity is reduced algebraically \cite{Koch2007}. This fact makes it possible to find a regime where charge noise is suppressed while sufficient anharmonicity is retained. Typical parameter choice for a transmon is $E_J/E_C \approx 100$ which results in the relative anharmonicity of few percent (specifically, one finds $\alpha_r=-0.04$ for $E_J/E_C= 113$). Increasing the anharmonicity is desirable if one wants to achieve higher fidelity of qubits and faster gate operations.
\begin{figure}[t]
	\centering
	\includegraphics[width=\columnwidth]{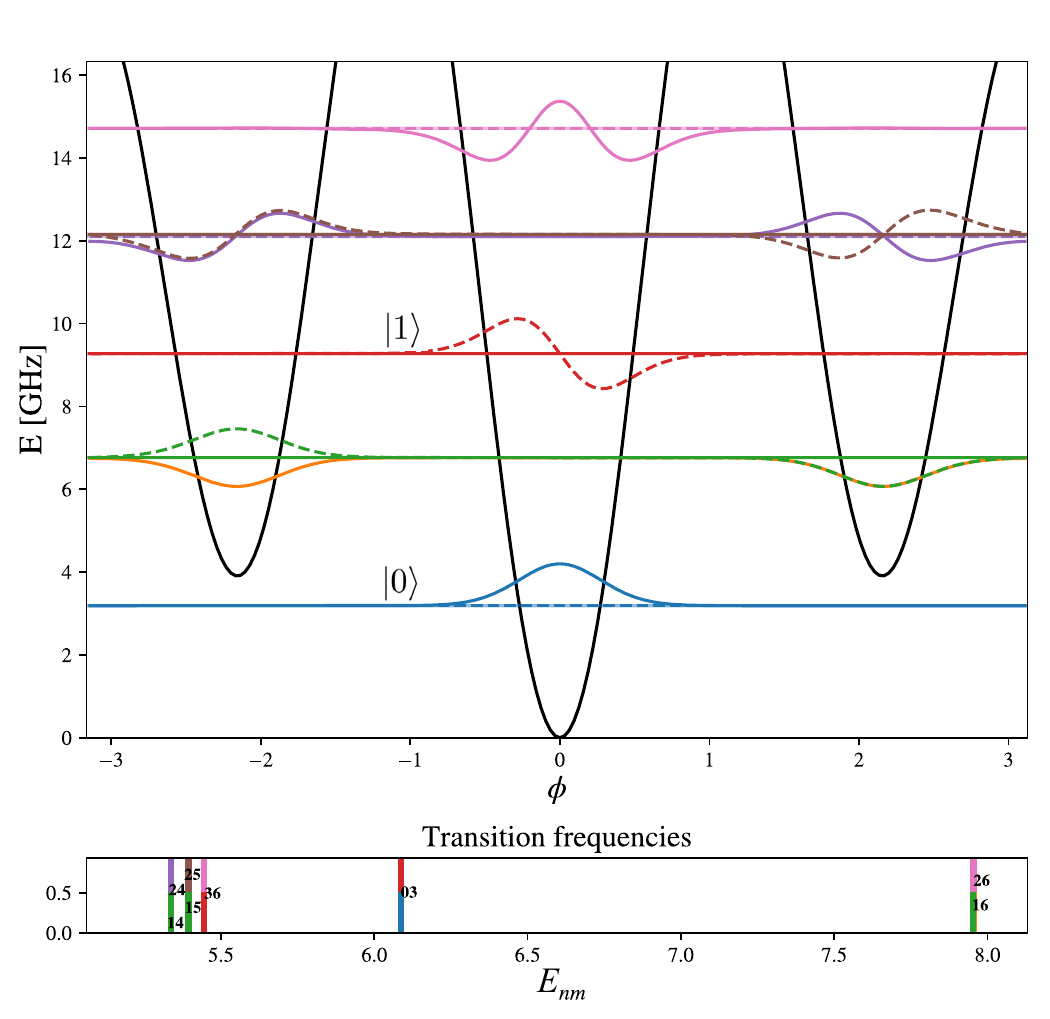}
	\caption{Top panel: Spectrum for $N=2$, $M=3$ NMon for $\beta=75$, $\eta=15$ with $\varphi_{\rm ext}=0$. ($E_N^J=4.5$GHz, $E_M^J=0.9$GHz, $E^C=0.06$GHz) Code space is defined as $n=0 \rightarrow |0\rangle$ and $n=3 \rightarrow |1\rangle$. Model parameters can be chosen (by overall rescaling of all energies) such that the qubit frequency $\omega_{01} = E_3-E_0 = 6.08$GHz which is a typical transmon frequency. In terms of frequency, closest transition from the excited state $|1\rangle$ is to $n=6$. For these parameters the relative anharmonicity is found to be $\alpha_r=-{0.105}$. Note that this is larger than typical relative anharmonicity of transmon regime where $-\alpha_r \approx 0.03-0.05$ corresponding to $200-300$MHz detuning from harmonic oscillator with energy spacing of $6$GHz. Bottom panel: Transition frequencies near $\omega_{01}$ ($E_{03}$) including all transitions between levels shown in the top panel (up to $n=6$) }
	\label{fig:23on_75_15_zeroflux}
\end{figure}

The key idea is to modify the Josephson potential to increase nonlinearity, which in turn can increase anharmonicity. The NMon potential consists of two cosine functions as indicated in Eq.\ \eqref{eq:nmon_intro}. For our subsequent discussion it is more convenient to start with a more general Hamiltonian Eq. \eqref{eq:nmonwithalpha} derived in Appendix A and choose $\alpha=M$. This results in 
\begin{multline}
\label{eq:32monwit}
H = 4E_Cn^2 -NE_J^N\cos \left(\phi -\kappa \frac{\varphi_{\rm ext}}{N}\right) \\ -ME_J^M\cos \left(\frac{N}{M}\phi  + (1-\kappa)\frac{\varphi_{\rm ext}}{M}\right) 
\end{multline} 
For simplicity, we focus in the following on the regime where $\varphi_{\rm ext}=0$ though we will still consider flux fluctuations. If we ignore the second term (set $E_J^M = 0$) the expression above is identical to the transmon or Cooper pair box Hamiltonian \cite{Nakamura1999}. Note also that the split transmon is a special case of NMon for $N,M=1$.
\begin{figure*}[t!]{
    \includegraphics[width=\textwidth]{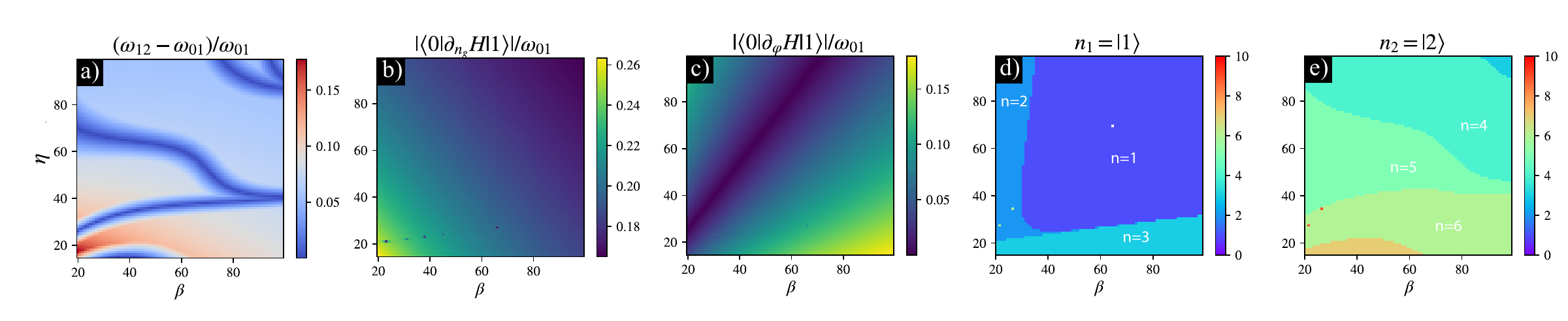}
\caption{Phase diagram ($N=2$, $M=3$) for a range of $E_J^N$ and $E_J^M$ in dimensionless parameters $\beta =E_J^N/E_C$ and $\eta =E_J^M/E_C$ at $\varphi_{\rm ext}=0$ and $n_g=0$ for $\kappa=0.5$ a) absolute value of the relative anharmonicity $|\alpha_r|$. b) renormalized transition matrix element for charge fluctuations c) matrix element for flux fluctuations. Note that this is the only plot that depends on the value of $\kappa$ for $\varphi_{\rm ext}=0$. Decreasing $0<\kappa<1$ rotates the line of zero clockwise. (See Fig. \ref{fig:23MontuningKlineofzeros}) This means the flux matrix elements for a point of interest in the parameter space can be made arbitrarily small by tuning the ratio of shunt capacitances.  d,e) lowest two energy levels above the ground state relevant for computational space $|0\rangle,|1\rangle$ and information loss $|1\rangle \rightarrow |2\rangle$ due to leakage outside the computational space. State $|1\rangle$ ($n=n_1$) is chosen such that it is a low energy excited state with the highest matrix element with the ground state $|0\rangle$ which is always $n=0$. State $|2\rangle$ ($n=n_2$) is chosen such that $|E_{n_2}-E_{n_1}|$ is closest to the qubit frequency $\omega_{01}=E_{n_1}-E_0$, therefore it is the state that is most relevant for information loss when the qubit is driven externally. }
\label{fig:23phasediagram_zeroflux}
}
\end{figure*}

Introducing another harmonic to the Josephson potential in Eq.~\eqref{eq:32monwit} comes at the price of spurious energy levels with wavefunctions localized in the additional potential wells which are outside the computational space, as illustrated in Fig.\ \ref{fig:23on_75_15_zeroflux}. We argue that these levels can be ignored because the transition matrix elements between the states localized in different potential wells are negligible compared to the ones within the computational space -- at least three orders of magnitude smaller for our specific example discussed  below.

For the parameters chosen in Fig.~\ref{fig:23on_75_15_zeroflux} we have $E_J^N=4.5$GHz,  $E_J^M=0.9$GHz and $E_C=60$MHz. The Hamiltonian in this case can me written as $H=4E_Cn^2 -2E_J^N\cos(3\phi)-3E_J^M\cos(2\phi)$. Therefore, if we only consider the leading $E_J^N$ term we have $2E_J^N/E_C \approx 150$ which is well into the transmon regime. For a typical transmon with $E_J=22.5$GHz and $E_C=200$MHz we find $E_J^N/E_C \approx 113$. In this sense, while reducing charge fluctuations NMon compromises less on anharmonicity. 

\begin{figure}[b!]
	\centering
	\includegraphics[width=\columnwidth]{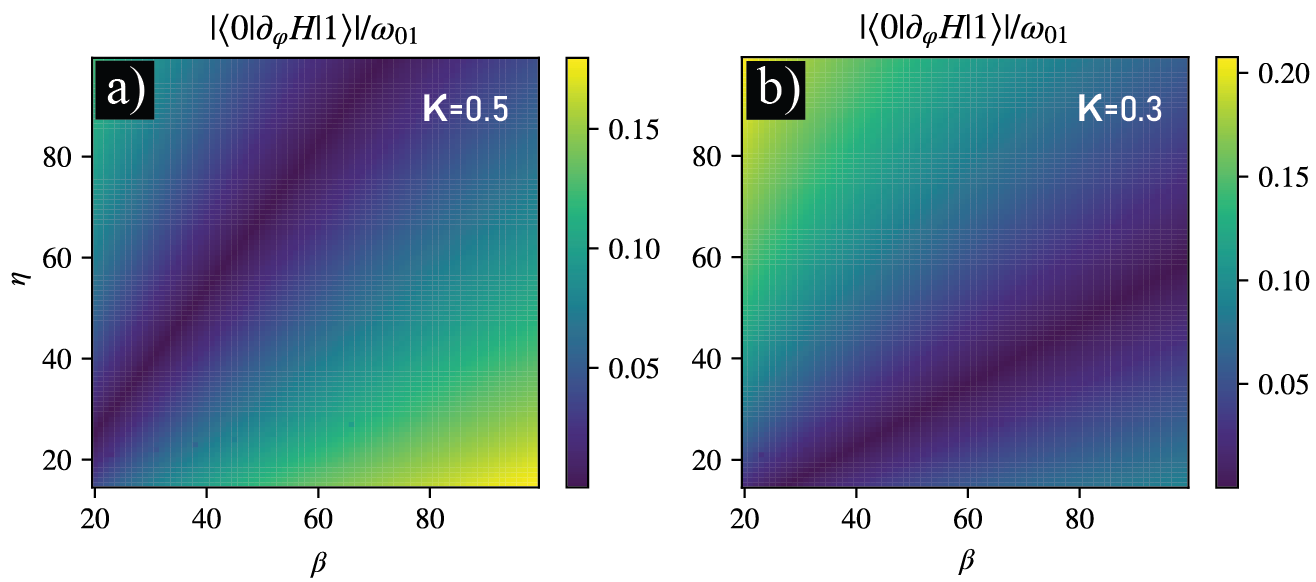}
	\caption{Flux fluctuation matrix elements for the $N=2$, $M=3$ NMon parameter space shown in Fig.~\ref{fig:23phasediagram_zeroflux}. Note that by picking different values of $\kappa$ we observe that the line where the matrix elements vanish can be moved. $\beta=E_J^N/E_C$ and $\eta=E_J^M/E_C$}
	\label{fig:23MontuningKlineofzeros}
\end{figure}

To confirm these expectations, we consider transition matrix elements that are scaled by the qubit frequency, which we define as the energy splitting between the computational states. This is a universal metric for comparing different qubit architectures since we do not have to match the gap to compare transition rates for the same type of noise channel. This can be regarded as a unit system where all energies are expressed in units of the qubit frequency, which can be tuned by rescaling all the energies in the system. However, the renormalized matrix elements remain invariant under such rescaling.

\begin{figure}[t!]
	\centering
	\includegraphics[width=\columnwidth]{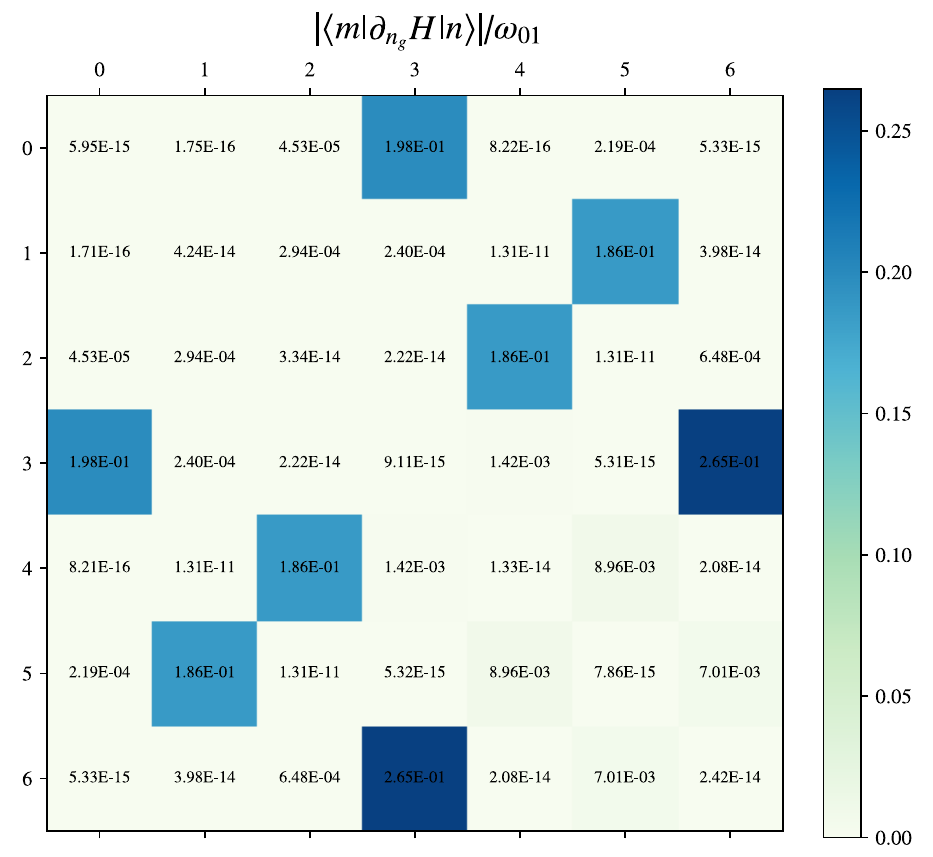}
	\caption{Charge fluctuation matrix elements for $N=2$, $M=3$ NMon for $\beta=75$, $\eta=15$ with $\varphi_{\rm ext}=0$. Code space is defined as $n=0 \rightarrow |0\rangle$ and $n=3 \rightarrow |1\rangle$ and this choice is dictated by the nature of the matrix elements: Note that the most significant matrix element that the ground state has with the excited states is $\langle 3 |\partial _{n_g} H | 0\rangle $ (between physical states) which is larger than any other matrix element (in the energetic vicinity of $|0\rangle$) by at least three orders of magnitude. Similarly, once the system is excited to $n=3$, the two most likely transitions are $3\rightarrow 0$ (back to the ground state) and $3\rightarrow 6$ (information leaking out of the code space)}
	\label{fig:23on_75_15_chargefluctuations}
\end{figure}

We compute the transition matrix elements due to charge and flux fluctuations. Details of these calculations can be found in Appendix C. These matrix element amplitudes (renormalized by the qubit frequency $\omega_{01}$) are used for estimation of relaxation times. The charge fluctuation matrix elements (in code space) $|\langle 0 | \partial_{n_g} H | 1 \rangle |/\omega_{01}= 0.2$ and $|\langle 1 | \partial_{n_g} H | 2 \rangle |/\omega_{01}= 0.27$ (Fig.~\ref{fig:23on_75_15_chargefluctuations}) are smaller (same order of magnitude) than any value in the vicinity of the typical transmon parameter regime (See also Fig.~\ref{fig:splittransmonphasediagram}). Flux matrix elements are found to be  $|\langle 0 | \partial_\varphi H | 1 \rangle |/\omega_{01}= 0.16$ and $|\langle 1 | \partial_\varphi H | 2 \rangle |/\omega_{01}= 0.18$ (taken from Fig.\ \ref{fig:appendix_combined}d in the Appendix) and are also comparable to what we find for split transmon (Fig.~\ref{fig:splittransmonphasediagram}) depending on the junction asymmetry parameter $d=E_J^N-E_J^M$ defined in Appendix E. Note that in the limit of perfect symmetry ($d\rightarrow 0$) between two junctions of the split transmon, the matrix elements vanish on a horizontal line in the parameter space shown in Fig.~\ref{fig:splittransmonphasediagram}.c and ~\ref{fig:splittransmonphasediagram}.f). We will demonstrate similar vanishing of matrix elements for NMon in the next section.  We note that this line can be `rotated' (see Fig.~\ref{fig:23MontuningKlineofzeros}) by tuning $\kappa$ such that flux matrix elements can be made arbitrarily small by moving this line closer to the chosen $(\beta,\eta)$ values in the phase diagram.

We therefore conclude that matrix elements of NMon can be tuned to be comparable to those of typical transmon values while having enhanced relative anharmonicity, with at least a factor of two in the specific case studied in Fig.~\ref{fig:23on_75_15_zeroflux}, as compared to the typical optimum value $\alpha_r=-0.04$ for transmon. 

\begin{figure}[b!]
	\centering
	\includegraphics[width=\columnwidth]{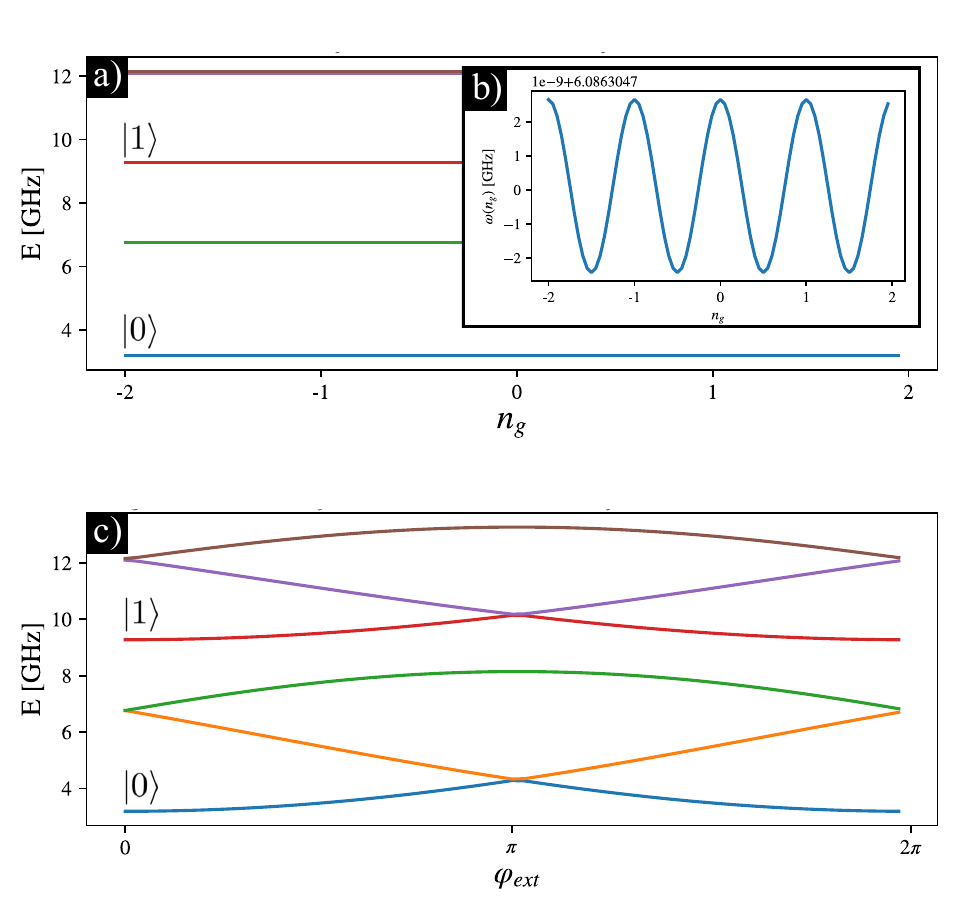}\caption{Spectrum for $N=2$, $M=3$ NMon for $\beta=75$, $\eta=15$ \textbf{a)} as a function of charge offset $n_g \in [-2,2)$ and \textbf{c)} as a function of the external flux $\varphi_{\rm ext} \in [0,2\pi)$ through the circuit loop. Model parameters chosen to be the same as Fig. \ref{fig:23on_75_15_zeroflux} where $\varphi_{\rm ext}=0$ is shown. The spectrum is virtually independent of $n_g$ as seen in panel \textbf{a)}. Inset panel \textbf{b)} shows that relative fluctuations in the qubit frequency are of the order $\Delta\omega/\omega \approx 10^{-9}$ as a function of offset charge. Note that the "sweet spot" external flux value is found to be $\varphi_{\rm ext}=0$ in panel \textbf{c)} where the band dispersion for the energy levels in the computational space have zero dispersion.}
	\label{fig:23on_75_15_tuneext}
\end{figure}
Next, we compute the dispersion of the energy levels as a function of the external flux $\varphi_{\rm ext}$ and offset charge $n_g$. Fig.~\ref{fig:23on_75_15_tuneext}.a confirms that the energy levels are virtually independent  of the offset charge (this means we are well into the transmon regime). Fig.~\ref{fig:23on_75_15_tuneext}.c shows that $\varphi_{\rm ext}=0$ is the ``sweet spot" in terms of minimizing dephasing due to the external flux fluctuations as the dispersion of the energy levels making up the computational space (blue-$|0\rangle$ and red-$|1\rangle$ curves) vanishes. 

In order to show that this behaviour is observed for a range of model parameters, we now explore the parameter space for a fixed external flux $\varphi_{\rm ext}=0$ and offset charge $n_g=0$, tune the Josephson energies $E_J^N$ ($\beta$) and $E_J^M$ ($\eta$) well into the transmon regime and compute anharmonicities and matrix elements while using an algorithm to determine the most suitable energy levels $|0\rangle$ ($n=0$), $|1\rangle$ ($n=n_1$) for a qubit. $|2\rangle$ ($n=n_2$) is determined as the energy level that the wavefunction in the computational space is most likely to leak out to. The algorithm and the phase space are described in Fig.~\ref{fig:23phasediagram_zeroflux}. Note that we find an extended region (red) in the phase space where the relative anharmonicity is enhanced to $\alpha_r \approx -0.1$ which is more than twice the typical transmon value. 

\section{Large Odd N, M=2}
The ideas presented in the previous section can be developed further by picking a large integer $N$ while keeping $M=2$ fixed. We will show that this configuration leads to smaller transition matrix elements (this would mean longer relaxation/dephasing times) as well as even larger anharmonicity values compared to what we found in the previous section.
 
\begin{figure}[t!]
	\centering
	\includegraphics[width=\columnwidth]{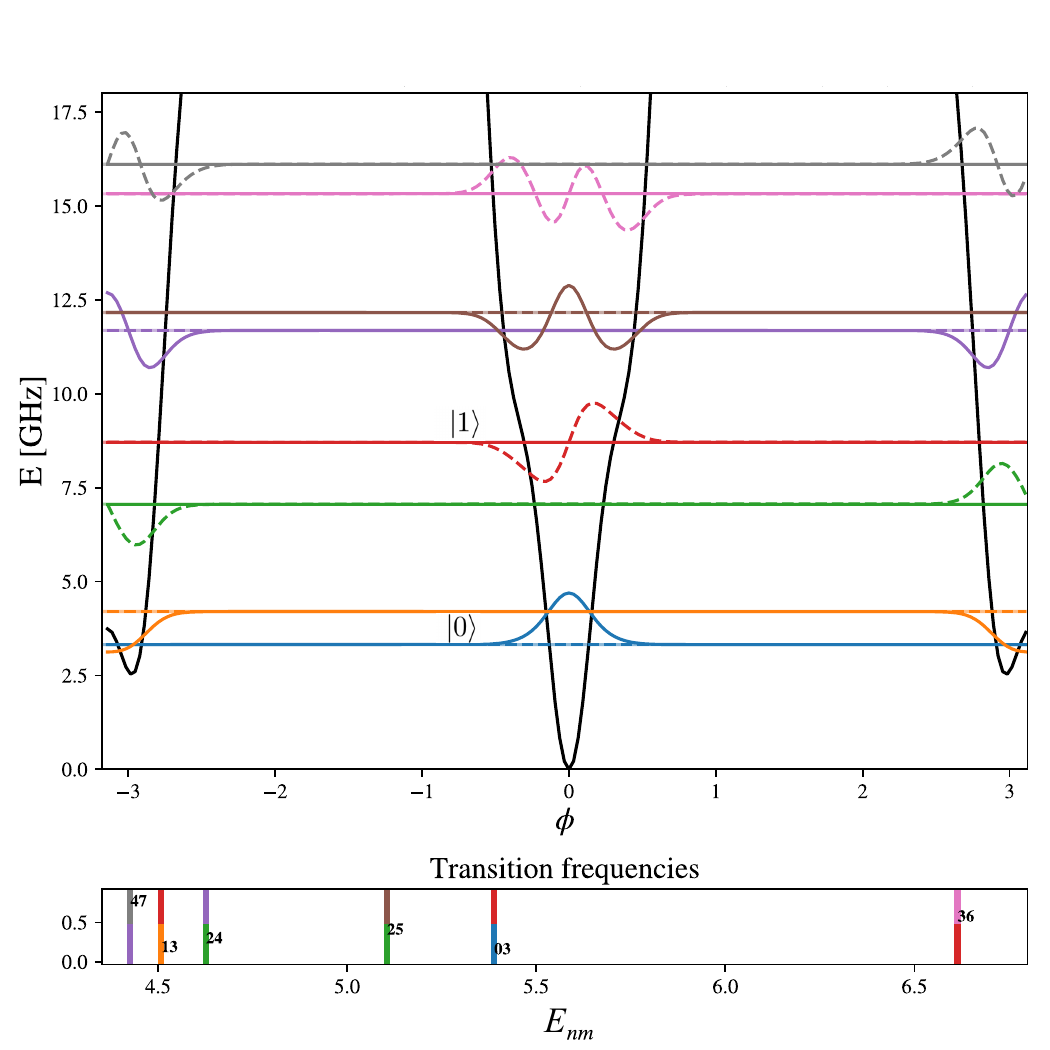}
	\caption{Top panel: Spectrum for $N=13$, $M=2$ NMon for $\beta=140$, $\eta=55$ with $\varphi_{\rm ext}=0$ and $\kappa=1$. ($E_N^J=4.5$GHz, $E_M^J=0.9$GHz, $E^C=0.06$GHz) Code space is defined as $n=0 \rightarrow |0\rangle$ and $n=3  \rightarrow |1\rangle$. Model parameters can be chosen (by overall rescaling of all energies) such that $\omega_{01} = 5.39$GHz which is a typical transmon frequency. Note that $n=5$ as the next excited state accessible from $n=3$. In this case the relative anharmonicity is found to be $\alpha_r=-0.359$.  Bottom panel: Transitions near $\omega_{01}$ including all transitions between levels shown in the top panel (up to $n=7$)}
	\label{fig:NMONdoublewell}
\end{figure}

\begin{figure}[h]
	\centering
	\includegraphics[width=\columnwidth]{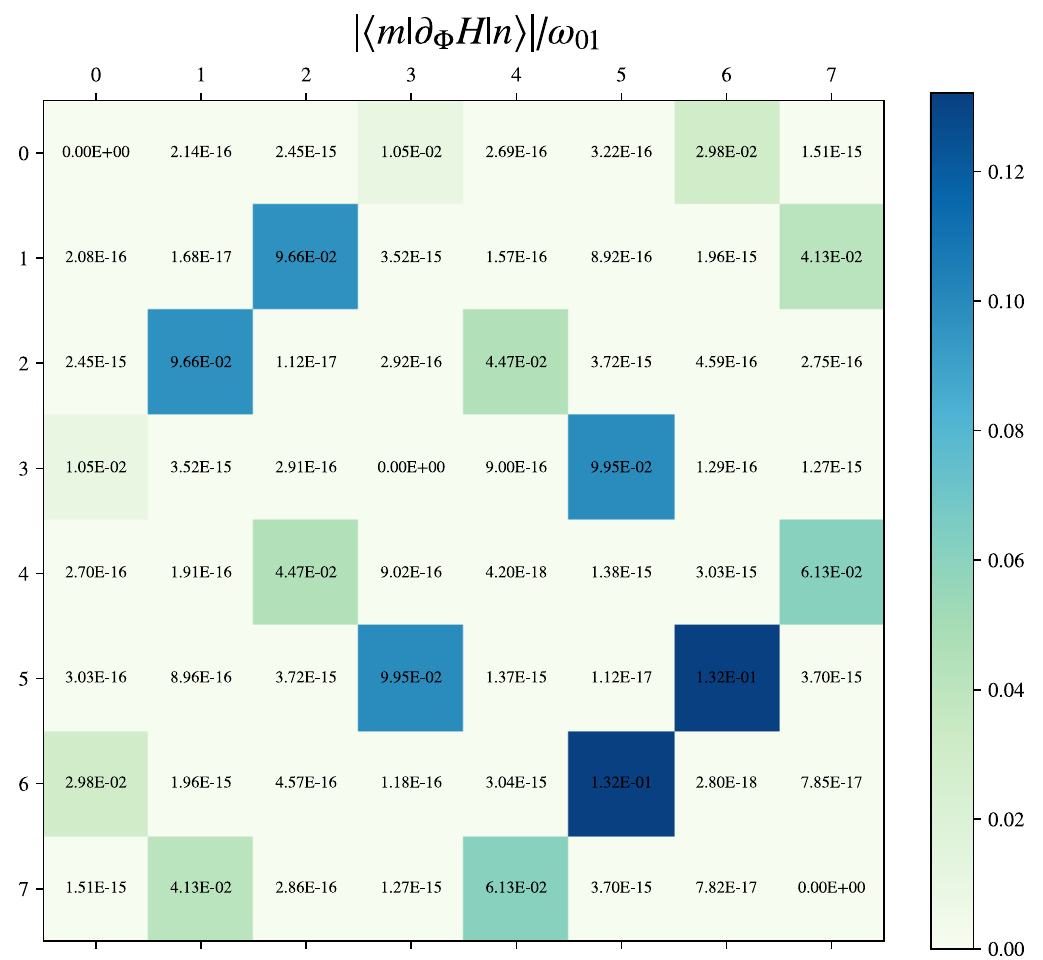}
	\caption{Flux fluctuation matrix elements between lowest 8 physical states for $N=13$, $M=2$ NMon for $\beta=140$, $\eta=55$ with $\varphi_{\rm ext}=0$. Code space is defined as $n=0 \rightarrow |0\rangle$ and $n=3 \rightarrow |1\rangle$ }
	\label{fig:132mon_140_55_fluxfluctuations}
\end{figure}

\begin{figure*}[t!]{
    \includegraphics[width=\textwidth]{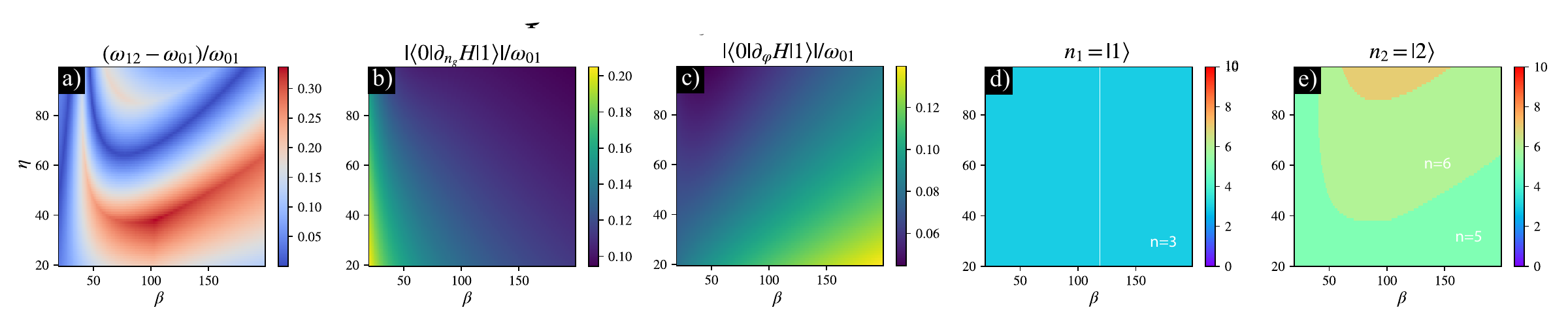}
	
\caption{Phase diagram for NMon for $N=13$ and $M=2$ for a range of $E_J^N$ and $E_J^M$ in dimensionless parameters $\beta =E_J^N/E_C$ and $\eta =E_J^M/E_C$ at zero flux $\varphi=0$ and $n_g=0$ for $\kappa=1$ Same quantities calculated as in Fig.~\ref{fig:23phasediagram_zeroflux}. Note that this configuration is able to achieve higher anharmonicity compared to transmon or the $N=2, M=3$ configuration we have discussed in the previous section. }
\label{fig:NMON_LARGEN_phasediagram}
}
\end{figure*}

For a concrete example, we consider the configuration with $N=13$ and $M=2$ leading to teh energy landscape depicted in Fig.~\ref{fig:NMONdoublewell}. This choice leads to a deep double well potential created by the $\cos{2\phi}$ term  modulated by a faster oscillating term $\cos{N\phi}$ which breaks the degeneracy between the energy levels of the double well pair. 

It is important that $N$ is an odd integer as the degeneracy is due to Cooper pair parity, which is broken when even and odd sectors are mixed since a term $\cos{k\phi}$ corresponds to a $k$ Cooper pair tunneling process. 
Despite introducing more junctions, this approach has the following advantages: the overlap between states localized in different wells is suppressed further by large-$N$ (suppressed tunneling due to deeper potential wells), which decouples the computational space even more compared to the $N=3,M=2$ case discussed above. This is in addition to even higher nonlinearity which is preferable for faster gate operations. Similar to fluxonium, the large-$N$ limit suppresses the flux fluctuation matrix elements, making it possible to control both charge and flux matrix elements by increasing $E_J$ and  $N$ respectively.

We compute the transition matrix elements for the parameters used in Fig.~\ref{fig:NMONdoublewell}. Flux matrix elements are found to be $|\langle 0 | \partial_\varphi H | 1 \rangle |/\omega_{01}= 0.01$ and  $|\langle 1 | \partial_\varphi H | 2 \rangle |/\omega_{01}= 0.10$ taken from Fig.~\ref{fig:132mon_140_55_fluxfluctuations} and $|\langle 0 | \partial_{n_g} H | 1 \rangle |/\omega_{01}= 0.11$ and $|\langle 1 | \partial_{n_g} H | 2 \rangle |/\omega_{01}= 0.12$ taken from Fig.~\ref{fig:appendix_combined}c. Comparing these values to fluxonium  ($|\langle 0 | \partial_\varphi H | 1 \rangle |/\omega_{01}= 6.9$ and $|\langle 0 | \partial_{n_g} H | 1 \rangle |/\omega_{01}= 0.13$) we note that while the charge matrix elements are comparable (see Appendix F for a discussion of fluxonium), the effect of flux fluctuations is suppressed by an order of magnitude, potentially leading to longer relaxation and dephasing times.

In terms of frequencies, closest transition from the excited state $|1\rangle$ is to $n=6$. However, the matrix elements for the transition $3 \rightarrow 6$ is extremely small ($\approx 10^{-16}$). Therefore we consider $n=5$ as the next excited state accessible from $n=3$. In this case the relative anharmonicity is found to be as large as $\alpha_r=-{0.359}$. (For $n=6$ it would be $\alpha_r={0.23}$)

Finally, we present the anharmonicity and transition matrix elements for a range of parameters in Fig.~\ref{fig:NMON_LARGEN_phasediagram}. Note that it is possible to achieve even larger anharmonicity in this configuration compared to the $N=2,M=3$ setup we studied above. 

\section{Population dynamics with flux drive} 

The two-level system we are considering can be manipulated by driving an oscillating flux through the circuit loop with the frequency of the desired transition, which is the qubit frequency $\omega_{03}$ in the specific example we have been considering for the parameters in Fig.~\ref{fig:NMONdoublewell}. One can solve the time dependent Schr\"odinger equation numerically to study the response of the qubit to flux drive. Starting with the initial state $|\Psi_0\rangle=\sum_m c_m(t)|m\rangle $ (in the absence of perturbation) where $H_0|m\rangle = E_m |m\rangle$ the time dependent perturbation $V(t)$ leads to
\begin{equation}
    i\hbar \partial_t c_n(t) = \sum_{n,m} \langle n|V(t)|m\rangle e^{i(E_n-E_m)t/\hbar}c_m(t)
\end{equation} 
which we solve for $c_n(t)$ determining the time evolution of the initial state in the presence of the perturbation to obtain $|\Psi(t)\rangle=\sum_n c_n(t)e^{-iE_n t}|n\rangle $. For flux drive the perturbation we consider is given by
\begin{equation}
    V(t) = \partial_\Phi H\delta \Phi(t).
\end{equation}

\begin{figure}[h!]
	\centering
	\includegraphics[width=\columnwidth]{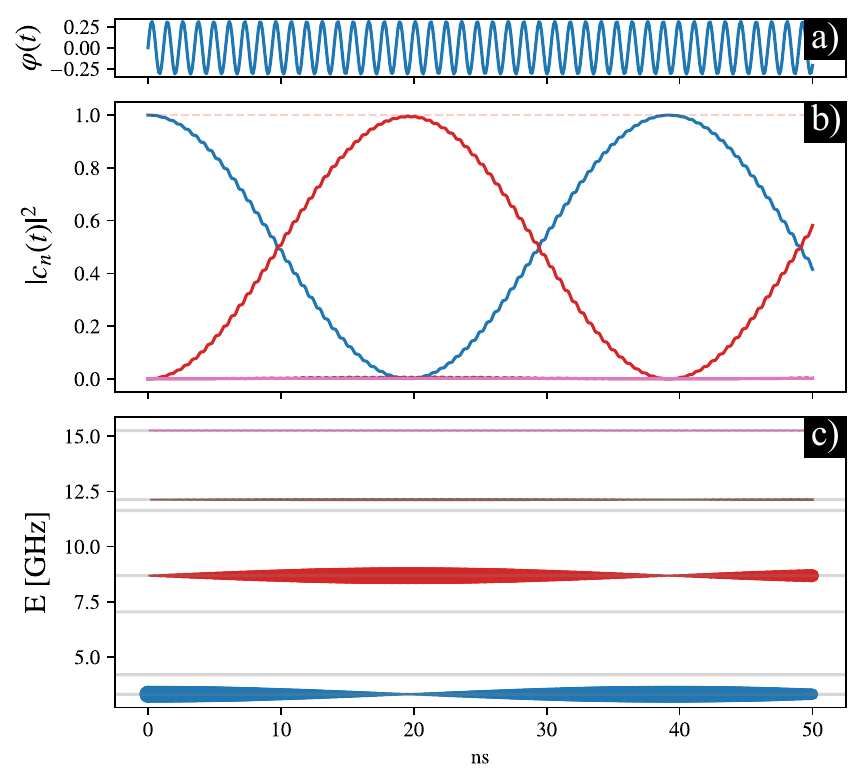} 
 \caption{Dynamics of system described in Fig.~\ref{fig:NMONdoublewell} for sinusoidal external flux drive. a) Flux drive in dimensionless flux $\varphi_{\rm ext}(t)$ as a function of time with amplitude $\approx 0.3$ and frequency $\omega_{01} = 5.39$GHz. b) Rabi oscillations as a result of the drive between $|0\rangle$ and $|1\rangle$ of the code space, demonstrating that state preparation is possible with minimal information leakage into other energy levels. c) Rabi oscillations shown in the context of population of the low-lying energy levels where the dot size for each time point is proportional to $|c_n(t)|^2$ for energy level $n$.}
	\label{fig:132mon_140_55_fluxdrive}
\end{figure}

In Fig.~\ref{fig:132mon_140_55_fluxdrive} we plot the solution with the initial condition that at $t=0$ the qubit is in the ground state $|0\rangle$. In order to observe complete population inversion, flux drive amplitude must be above a certain threshold which is around $|\varphi_{\rm ext}|\approx 0.18$. We pick $\varphi_{\rm ext}= 0.3$  and observe that increasing this value increases the Rabi oscillation frequency.

\section{Summary and Conclusion} 

We have shown that the NMon circuit geometry is capable of exhibiting twice as much anharmonicity (for $N=2$, $M=3$) comparable to existing transmon architectures while having comparable  transition matrix elements. We have also demonstrated how this architecture can be improved by considering arrays with large number of junctions, making NMon in this regime an alternative to fluxonium due to its even lower susceptibility to flux noise. Our results suggest that large-$N$ configurations of the type we have studied in Sec.\ III might outperform fluxonium in terms of relaxation time due to an improvement by an order of magnitude in terms of flux transition matrix elements. 

All our results are based on assuming that junctions within the same arm are identical. However, fabrication imperfections will inevitably lead to a distribution of Josephson energies and capacitances. Such imperfections are unavoidable and have been treated perturbatively \cite{Ferguson2013} for the case of fluxonium  (corresponding to NMon with $M=1$ and $N \gg 1$), justifying the superinductance mode approximation. Similar calculations can be carried out for the more general case of NMon, which we leave for future work.

Having explored only a small part of the configuration space, we demonstrated that further improvements are possible for the existing qubit architectures. One can also consider a flux-qubit type scenario in this system, choosing the lowest lying two levels in Fig.~\ref{fig:NMONdoublewell} as the code space where the two wavefunctions are localized in separate wells. Note that the matrix elements between the two are extremely suppressed (see 0-1 matrix elements in Figs.~\ref{fig:132mon_140_55_fluxfluctuations}, \ \ref{fig:appendix_combined}c) making direct transitions between these two states very difficult. Nevertheless, it may be possible to construct schemes where transitions between the code space energy levels are achieved through higher lying intermediary states.

\label{sec:conclusion}

\section{Acknowledgments}
The authors are indebted to Vedangi Pathak, Timothy Duty, Joe Salfi and Alexandre Blais for inspiring discussions and correspondence. This work was supported by NSERC, CIFAR and the Canada First Research Excellence Fund, Quantum Materials and Future Technologies Program.

\newpage 
\bibliography{main}

\section{Appendix}

\setcounter{figure}{0}
\setcounter{table}{0}
\setcounter{equation}{0}
\makeatletter
\renewcommand{\theequation}{A\arabic{equation}}
\renewcommand{\thefigure}{A\arabic{figure}}

\subsection{Quantizing the NMon for time dependent flux  using irrotational constraint}

We assume the junctions in each arm are identical - and therefore constrained in a way that there is only one phase-flux variable describing all of them. The Lagrangian of the circuit is given by  
$ \mathcal{L} = T - U$ where $$T = \frac{1}{2}MC_M \dot{\Phi}_M^2 + \frac{1}{2}NC_N \dot{\Phi}_N^2$$  
and
$$ U = -NE_J^N\cos(2\pi\Phi_N/\Phi_0) -ME_J^M\cos(2\pi\Phi_M/\Phi_0) $$ 
where we identify a generalized flux "coordinate" $\Phi_\alpha$ for each junction such that $2\pi\Phi_\alpha/\Phi_0 = \phi_\alpha$ where $\alpha=M,N$. Continuity of the order parameter implies the constraint
$$ \Phi_{\rm ext} = M\Phi_M - N\Phi_N$$ 
We define a new effective coordinate as a linear combination of the original flux variables
$$ \Phi =  \alpha\Phi_M + \beta\Phi_N. $$ 
These two equations can be inverted and yield
$$ \Phi_M = \frac{N\Phi +\beta\Phi_{\rm ext}}{\alpha N + \beta M},   \hspace{15pt} \Phi_N = \frac{M\Phi -\alpha\Phi_{\rm ext}}{\alpha N + \beta M}.$$ 
With these expressions, we can write the Lagrangian $\mathcal{L}$ in the new variables  and compute the conjugate momenta $Q_\Phi = \partial \mathcal{L}/\partial \dot{\Phi}$ and $Q_{\Phi_{\rm ext}} = \partial \mathcal{L}/\partial \dot{\Phi}_{\rm ext}$ which we need for the Legendre transform to obtain the Hamiltonian
$$H = Q_{\Phi}\dot{\Phi} + Q_{\Phi_{\rm ext}}\dot{\Phi}_{\rm ext} - T + U.$$
This becomes
\begin{multline}
H = \frac{NM(NC_M + MC_N)}{2(\alpha N +\beta M)^2}\dot{\Phi}^2  \\ + \frac{NM(\beta C_M - \alpha C_N)}{(\alpha N +\beta M)^2}\dot{\Phi}\dot{\Phi}_{\rm ext}  \\
+  \frac{\beta^2 M C_M + \alpha^2 N C_N}{2(\alpha N +\beta M)^2}\dot{\Phi}^2_{\rm ext} + U(\Phi,\Phi_{\rm ext}),
\end{multline} 
 where 
 \begin{multline}
  U(\Phi,\Phi_{\rm ext}) = -NE_J^N\cos \left(\frac{2\pi}{\Phi_0}\left[\frac{M\Phi -\alpha\Phi_{\rm ext}}{\alpha N + \beta M}\right]\right) 
  \\ -ME_J^M\cos \left(\frac{2\pi}{\Phi_0}\left[\frac{N\Phi +\beta\Phi_{\rm ext}}{\alpha N + \beta M}\right]\right). 
  \end{multline} 
We pick the new coordinate transformation coefficients $\alpha,\beta$ obeying the irrotational constraint \cite{You2019} such that the second term proportional to $\dot{\Phi}_{\rm ext}$ vanishes: 
$$ \beta = \frac{C_N}{C_M}\alpha.$$
We also drop the term proportional to $\dot{\Phi}_{\rm ext}^2$ as it will be a constant offset in the quantum mechanical problem when we quantize $\Phi$ only. Rewriting the Hamiltonian in these new variables as well as the conjugate momentum 
$$
Q_{\Phi} = \frac{NM(NC_M + MC_N)}{(\alpha N +\beta M)^2}\dot{\Phi} \rightarrow 2e\hat{n}
$$
and rescaling 
$$
\alpha \rightarrow \frac{C_M}{C_M N + C_N M}\alpha$$
(which implies   $\alpha N + \beta M \rightarrow \alpha$)
 we find that  
\begin{multline}
H = \frac{2e^2\hat{n}^2\alpha^2}{NM(C_M N + C_N M)} 
\\ -NE_J^N\cos \left(\frac{M\phi}{\alpha} -\frac{C_M \varphi_{\rm ext}}{C_M N + C_N M}\right) 
\\ -ME_J^M\cos \left(\frac{N\phi}{\alpha} +\frac{C_N \varphi_{\rm ext}}{C_M N + C_N M}\right)  \end{multline}
where $\alpha$ can be any finite value which we can pick $\alpha=1$ for convenience. Note that we have also converted flux variables to phase variables using 
$$
\phi = \frac{2\pi\Phi}{\Phi_0}, \hspace{10pt} \varphi_{\rm ext} = \frac{2\pi\Phi_{\rm ext}}{\Phi_0}.
$$

We can also consider the case where each junction array is shunted by a capacitor with capacitances $C_S^N$ and $C_S^M$, respectively. We can show that in our calculation above, we simply need to make the replacements  $C_N \rightarrow C_N + NC_S^N$ and $C_M \rightarrow C_M + MC_S^M$. This comes from adding the capacitance energies with volt-
age matching constraint across parallel $N (M)$ junctions.
In the limit where shunt capacitances are much larger than the junction capacitances, which is conducive for reaching the transmon limit $E_C \ll E_J$ where charge noise is suppressed, we simply need to replace $ C_N \rightarrow  NC_S^N$ and $ C_M \rightarrow  MC_S^M$. This large shunt capacitance limit simplifies the Hamiltonian
\begin{multline} \label{eq:nmonwithalpha}
H = \frac{2e^2\hat{n}^2\alpha^2}{(NM)^2(C_S^M + C_S^N)} \\ -NE_J^N\cos \left(\frac{M\phi}{\alpha} -\frac{C_S^M}{C_S^M + C_S^N}\frac{\varphi_{\rm ext}}{N}\right) \\ -ME_J^M\cos \left(\frac{N\phi}{\alpha}  + \frac{C_S^N}{C_S^M + C_S^N}\frac{\varphi_{\rm ext}}{M}\right)
\end{multline}.
Let us pick $\alpha=1$ and define $0 < \kappa < 1$
$$
\kappa = \frac{C_S^M}{C_S^N+C_S^M}
$$ 
which is a parameter controlled by the ratio of the shunt capacitances. Hamiltonian then becomes
\begin{multline}
H = 4E_C\hat{n}^2 -NE_J^N\cos \left(M\phi -\kappa \frac{\varphi_{\rm ext}}{N}\right) \\  -ME_J^M\cos \left(N\phi  + (1-\kappa)\frac{\varphi_{\rm ext}}{M}\right),
\end{multline}
where
$$  E_C = \frac{2e^2}{4(NM)^2(C_S^M + C_S^N)}. $$

\subsection{Numerical approach} \label{sec:numerical}
Starting with the Hamiltonian in the phase representation, we switch to the charge representation which is more tractable for exact diagonalization. Given the Hamiltonian of the form $$ H = 4E_C(-i\partial_\phi + n_g)^2 - E_J\cos(k\phi+\varphi)$$ where $k$ is an integer and $\varphi=\varphi_{\rm ext}$ represents dimensionless flux or a phase shift. NMon Hamiltonian above consists of these two types of terms only. We define the conjugate variable, number of Cooper pairs $n$, which we can switch to by defining $$|n\rangle = \frac{1}{\sqrt{N_0}}\sum_\phi e^{-in\phi} |\phi\rangle$$ where $N_0$ is the number of points in our discretization scheme. We then compute the matrix elements $\langle n | H | m \rangle$ in the charge representation
\begin{multline*}
\langle n | H | m \rangle =  4E_C(n + n_g)^2\delta_{n,m} \\ - \frac{E_J}{2}e^{i\varphi}\delta_{n,m-k} - \frac{E_J}{2}e^{-i\varphi}\delta_{n,m+k} \end{multline*}
We also need the matrix elements of the derivative of the Hamiltonian (in the eigenbasis of $H_0$, which we will call $|u\rangle$) with respect to external parameters, which we can compute in the following way. Consider the derivative of the matrix element which we have just computed (note the particle eigenstates do not depend on the external parameter $\lambda$)
\begin{align}\frac{\partial \langle n | H | m \rangle}{\partial \lambda} &= \langle n |\frac{\partial  H }{\partial \lambda}| m \rangle \\
& = \langle n | u \rangle \langle u | \frac{\partial  H }{\partial \lambda} | v \rangle \langle v | m \rangle \\ 
& = M_{nu} \langle u | \frac{\partial  H }{\partial \lambda} | v \rangle M^\dagger_{vm}
\end{align}
 where we defined $M_{mv}=\langle m | v \rangle$ which diagonalizes $M^\dagger H_0 M = E_0$. Inverting the last line, we obtain 
$$ 
\langle u | \partial_\lambda  H | v \rangle =  M^\dagger \frac{\partial}{\partial \lambda} \left( \langle n | H | m \rangle \right)  M
$$ where we know the expression $\langle n | H | m \rangle$ from above. We need to compute the derivatives given the fluctuating parameter and evaluate them using the expression above. 

\subsection{ Noise Calculations} \label{sec:melements_derivation}

External parameters (which we will call $\lambda$ for general discussion) of the Hamiltonian such as charge offset $n_g$ and external flux $\Phi_{\rm ext}$ fluctuate about the values we intend to set for optimal performance of a qubit. These classical variables therefore contribute to the decoherence in addition to other channels such as quasiparticle excitations or radiation. For small fluctuations of the classical parameter $\delta \lambda(t)$ around $\lambda_0$, we can expand $$ H(\lambda) = H(\lambda_0) + \frac{\partial H}{\partial \lambda}\delta \lambda(t). $$ 

We use time dependent perturbation theory to first order to find the transition amplitude $c_{ni}(t)=\langle n | U(t,0) |i\rangle$ that we will find the system in state $|n\rangle$ provided that the system is initially in state $|i\rangle$ before we turn on the perturbation $V(t)=\frac{\partial H}{\partial \lambda}\delta \lambda(t)$. We treat $H_0 = H(\lambda_0)$ as the unperturbed Hamiltonian satisfying $H_0|m\rangle = E_m |m\rangle$ and use the textbook result to first order 
$$ c_{ni}^{(1)}(t) = -\frac{i}{\hbar}\int_0^t e^{i\omega_{ni}t'} V_{ni}(t'), $$ 
where we defined $V_{ni}(t) = \langle n | V(t) | i \rangle$ and $\hbar\omega_{ni} = E_n - E_i$. The transition probability $p_{ni}(t) = |c_{ni}(t)|^2$ is then given by
        $$ p_{ni}(t) = \frac{|\langle n| \partial_\lambda H | i\rangle|^2}{\hbar^2} \int_0^t dt_1 \int_0^t dt_2 e^{i\omega_{ni}(t_1-t_2)} \delta \lambda(t_1)\delta \lambda(t_2).$$ 
        After a change of variables by defining $\tau=t_1-t_2$ and $T=(t_1+t_2)/2$ and time averaging we find
    \begin{multline*}
     p_{ni}(t) = \frac{|\langle n| \partial_\lambda H | i\rangle|^2}{\hbar^2} \int_0^t dT \\ \int_{-\infty}^{\infty} d\tau e^{i\omega_{ni}\tau} \langle\delta \lambda(T+\tau/2)\delta \lambda(T-\tau/2)\rangle,  \end{multline*} 
     where we set bounds on the $\tau$ integral  to infinity as appropriate  for small autocorrelation times \cite{Schoelkopf2002}
       
    If we further assume time translation invariance (stationary noise correlation function), then the $T$ integral becomes trivial. We thus have
    $$ p_{ni}(t) = \frac{|\langle n| \partial_\lambda H | i\rangle|^2}{\hbar^2} t S_\lambda(\omega_{ni})$$
    where we introduced the noise spectral density $$ S_\lambda(\omega) = \int_{-\infty}^{\infty} d\tau e^{i\omega\tau} \langle\delta \lambda(\tau)\delta \lambda(0)\rangle $$ which characterizes the time dependence of the fluctuations of the external parameter $\lambda$. The transition rate $\Gamma_{ni}$ is found by computing the time derivative of $p_{ni}$: $$\Gamma_{ni}= \frac{|\langle n| \partial_\lambda H | i\rangle|^2}{\hbar^2} S_\lambda(\omega_{ni}).$$

    As we have just seen, the transition rates are directly related to transition matrix elements due to external parameter perturbations. In this paper we will not be estimating relaxation/dephasing times. Instead we will study the transition matrix elements to compare performances of different qubit architectures.

\subsection{Split transmon} 
\label{sec:reviewingsplittransmon}
\begin{figure}[t]
	\centering
	\includegraphics[width=\columnwidth]{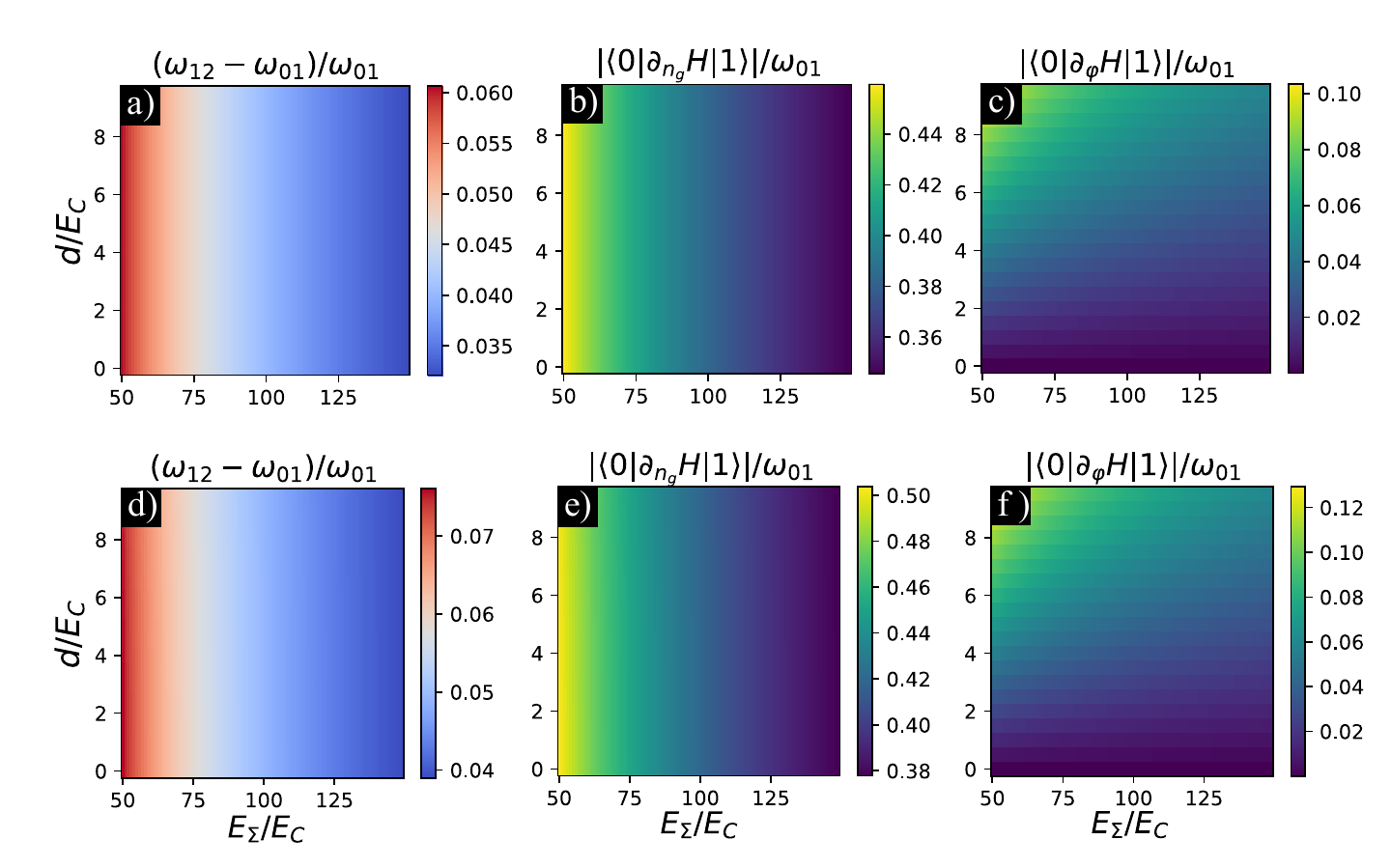}
	\caption{ Model parameter space for split transmon at $n_g=0$ and $\varphi_{ext}=0$ tuning $E_\Sigma$ and $d$ in units of charging energy $E_C$ showing a) relative anharmonicity b) charge fluctuation matrix elements c) flux fluctuation matrix elements for split transmon at $n_g=0$ and  $\varphi_{\rm ext}=0$. Panels d,e,f) show same quantities as a,b,c) except for $n_g=0$ and $\varphi_{\rm ext}=\pi$. }
	\label{fig:splittransmonphasediagram}
\end{figure}

\noindent NMon reduces to split transmon for parameters $N=1$ and $M=1$. Defining $E_\Sigma = E_J^N + E_J^M$ and asymmetry parameter $d=E_J^N-E_J^M$ (note that junction asymmetries typically lead to $d=\pm 10 \% $ \cite{Koch2007} in fabrication process).  Equation \eqref{eq:nmon_intro} reduces to
\begin{equation*}
 H = 4E_C\hat{n}^2 -E_\Sigma\cos\frac{\varphi_{\rm ext}}{2}\cos\phi - d\sin\frac{\varphi_{\rm ext}}{2}\sin\phi 
\end{equation*}
for $\kappa = 1/2$ which we can obtain by choosing equivalent shunt capacitances for both junctions. We present the transition matrix elements and relative anharmonicity for a range of model parameters in Fig.~\ref{fig:splittransmonphasediagram} for $\varphi_{\rm ext}=0$. A typical transmon \cite{McEwen2021} with transition frequency $\omega = 6$GHz can be achieved by picking $E_J=22.5$GHz and $E_C=200$MHz we find $E_J^N/E_C \approx 113$ where anharmonicity is $\alpha_r=-0.04$

\subsection{Fluxonium}\label{sec:fluxonium}

\noindent NMon reduces to the fluxonium qubit if we pick $M=1$ and $N=10$, a large number which leads to suppression of the flux fluctuations in addition to having giant anharmonicity. Starting with the master Hamiltonian Eq.  \eqref{eq:nmonwithalpha}, for $M=1$ we find that 
\begin{multline}
H = \frac{2e^2\hat{n}^2}{(C_S^M + C_S^N)} -NE_J^N\cos \left(\frac{\phi}{N} -\frac{C_S^M}{C_S^M + C_S^N}\frac{\varphi_{\rm ext}}{N}\right) \\ -E_J^M\cos \left(\phi  + \frac{C_S^N}{C_S^M + C_S^N}\varphi_{\rm ext}\right)
\end{multline}

To connect with the existing literature, we picked $\alpha=N$. In the large-$N$ limit, the second term generates the quadratic potential which is called the inductance term. Experimentally relevant case of inductor allocation \cite{Bryon2023} corresponds to $C_S^M \gg C_S^N$:
\begin{multline}
H = 4E_C\hat{n}^2  -NE_J^N\cos \left(\frac{\phi-\varphi_{\rm ext}}{N}\right) -E_J^M\cos \left(\phi\right)
\end{multline} where $E_C = e^2/2C_S^M$. This expression reduces to the familiar fluxonium Hamiltonian:
\begin{multline}
H = 4E_C\hat{n}^2  + \frac{E_L}{2}\left(\phi-\varphi_{\rm ext}\right)^2 -E_J^M\cos \phi
\end{multline} 

where we have dropped the constant term and defined $E_L=E_J^N/N$. We compute the matrix elements at $\varphi_{\rm ext}=\pi$ flux and $n_g=0$ offset charge in Fig.~\ref{fig:appendix_combined}a and Fig.~\ref{fig:appendix_combined}b. For these parameters, we find that $|\langle 0 | \partial_\varphi H | 1 \rangle |/\omega= 6.9$ and $|\langle 0 | \partial_{n_g} H | 1 \rangle |/\omega= 0.13$ 
\vspace{20pt}
\begin{figure}[t!]
	\centering
	\includegraphics[width=\columnwidth]{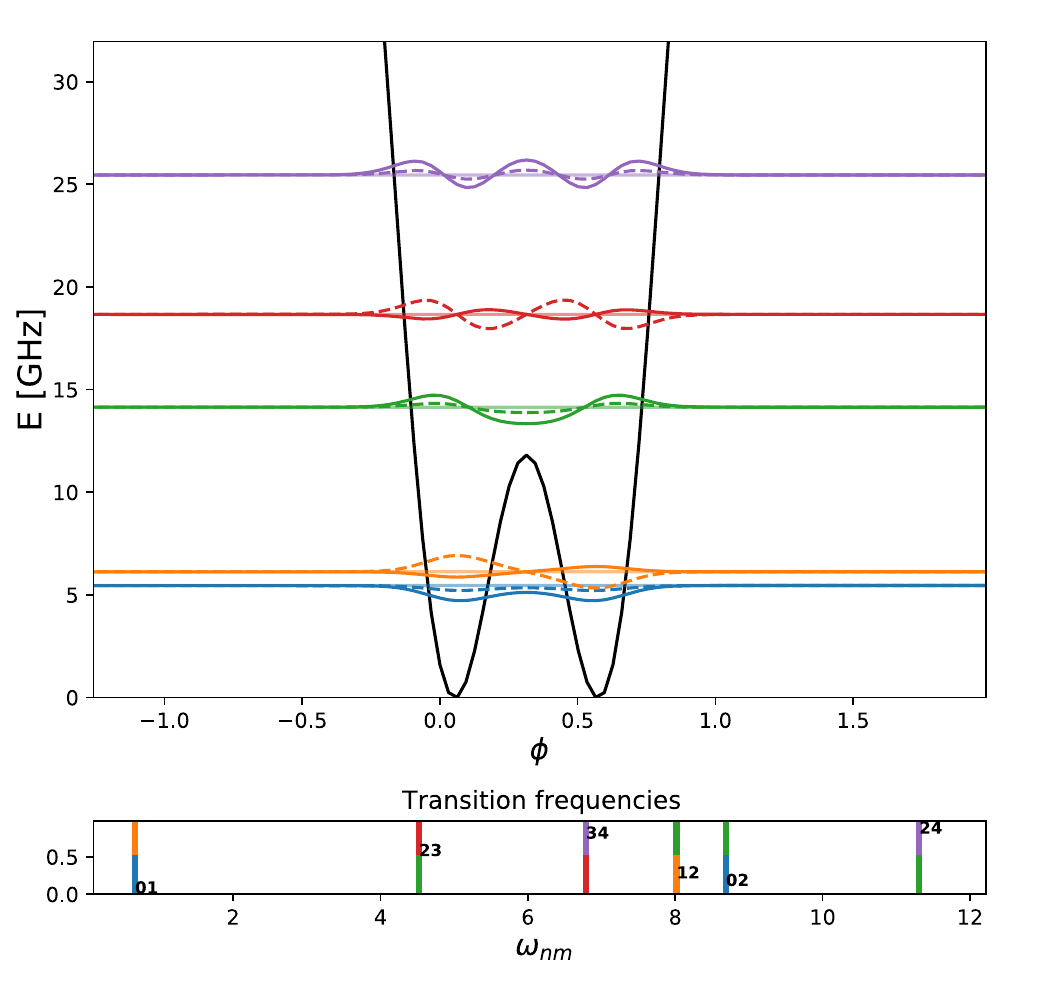}
	\caption{ Fluxonium spectrum for $N=10$, $M=1$ and $\kappa = 1$ ($E_N^J=4.5$GHz, $E_M^J=0.9$GHz, $E^C=0.06$GHz). At half flux quantum $\varphi_{\rm ext}=\pi$ the potential well develops two minima which leads to giant anharmonicity. }
	\label{fig:fluxoniumN10}
\end{figure}

\newpage

\begin{figure*}[t!]{
    \includegraphics[width=\textwidth]{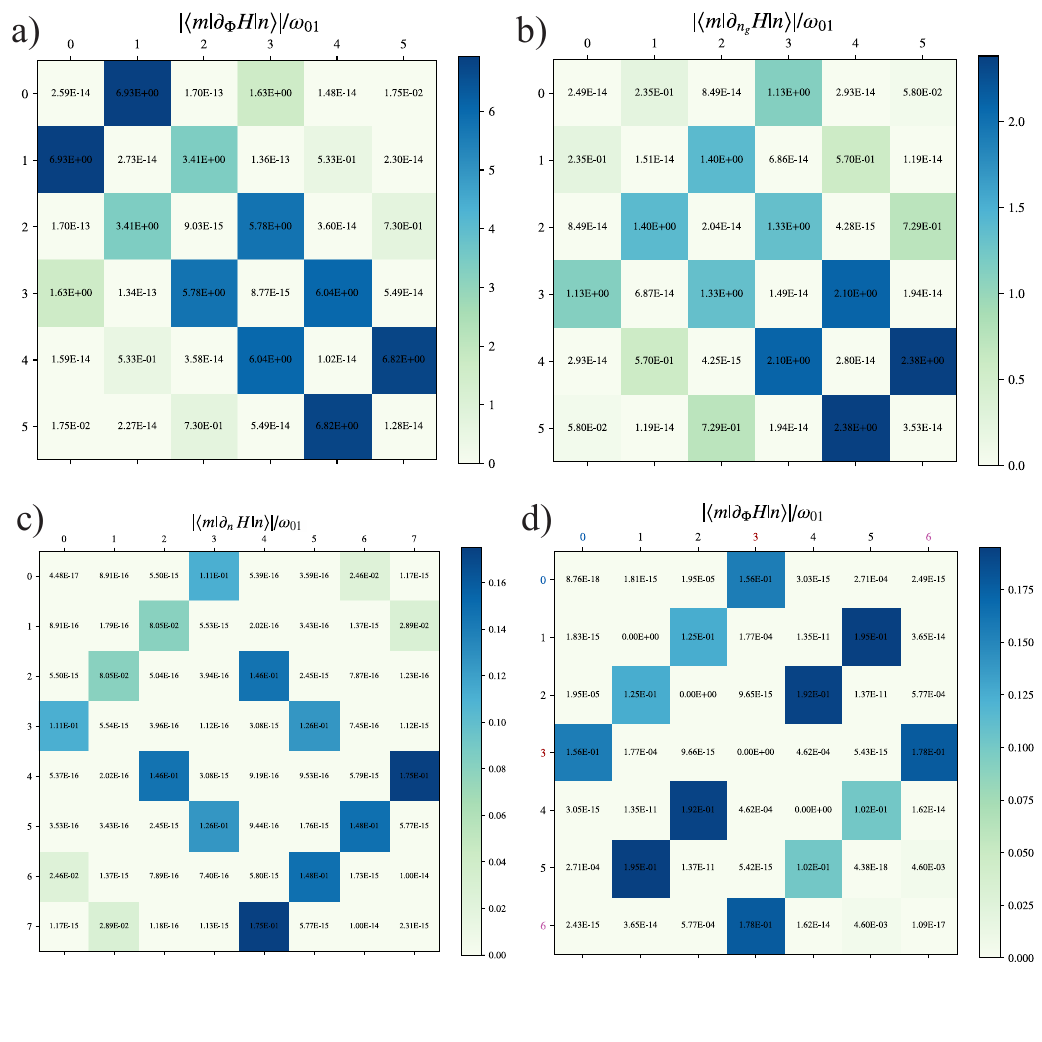}
    \caption{Panels \textbf{a)} and \textbf{b)} show the qubit frequency renormalized (See \ref{sec:melements_derivation}) matrix elements in the case of fluxonium with $N=10$ and $M=1$ for flux and charge fluctuations, respectively. Associated energy spectrum is shown in Fig. \ref{fig:fluxoniumN10}. Panel \textbf{c)} shows $N=13$, $M=2$ NMon charge fluctuation matrix elements computed for $\beta=75$, $\eta=15$ with $\varphi_{\rm ext}=0$. Code space is defined as $n=0 \rightarrow |0\rangle$ and $n=3 \rightarrow |1\rangle$. Panel \textbf{d)} shows $N=2$, $M=3$ NMon flux fluctuation matrix elements computed at $\beta=75$, $\eta=15$ with $\varphi_{\rm ext}=0$. Code space is defined as $n=0 \rightarrow |0\rangle$ and $n=3 \rightarrow |1\rangle$. }
\label{fig:appendix_combined}
}
\end{figure*}
 %


\end{document}